\documentclass[%
 preprint,
 amsmath,amssymb,
 aps, physrev,
]{revtex4-2}

\usepackage{graphicx}
\usepackage{dcolumn}
\usepackage{bm}

\begin{document}


\title{\textbf{Incorporating rank-free coupling and external field via an incoherent modulated spatial photonic Ising machine} 
}%

\author{Ze Zheng}
\author{Yuegang Li}
\author{Hang Xu}
\affiliation{State Key Laboratory of Photonics and Communications, Institute for Quantum Sensing and Information Processing, Shanghai Jiao Tong University, Shanghai 200240, P.R. China}
\affiliation{Hefei National Laboratory, Hefei, 230088, P.R. China}%

\author{Jingzheng Huang}
\affiliation{State Key Laboratory of Photonics and Communications, Institute for Quantum Sensing and Information Processing, Shanghai Jiao Tong University, Shanghai 200240, P.R. China}
\affiliation{Hefei National Laboratory, Hefei, 230088, P.R. China}
\affiliation{Shanghai Research Center for Quantum Sciences, Shanghai, 201315, P.R. China}

\author{Tailong Xiao}
 \email{tailong\_shaw@sjtu.edu.cn}
\affiliation{State Key Laboratory of Photonics and Communications, Institute for Quantum Sensing and Information Processing, Shanghai Jiao Tong University, Shanghai 200240, P.R. China}
\affiliation{Hefei National Laboratory, Hefei, 230088, P.R. China}
\affiliation{Shanghai Research Center for Quantum Sciences, Shanghai, 201315, P.R. China}

\author{Guihua Zeng}
 \email{ghzeng@sjtu.edu.cn}
\affiliation{State Key Laboratory of Photonics and Communications, Institute for Quantum Sensing and Information Processing, Shanghai Jiao Tong University, Shanghai 200240, P.R. China}
\affiliation{Hefei National Laboratory, Hefei, 230088, P.R. China}
\affiliation{Shanghai Research Center for Quantum Sciences, Shanghai, 201315, P.R. China}


\date{\today}

\begin{abstract}
Spatial photonic Ising machines offer a novel optical platform for optimization and spin-model simulation, but existing diffraction-based schemes rely on auxiliary spins or multiplexing to encode high-rank couplings and external fields, reducing either speed or spin count. We demonstrate an amplitude-only, rank-free spatial photonic Ising machine in which arbitrary Ising Hamiltonians are encoded as Hadamard products on aligned amplitude and binary spatial modulators and read out by a single-pixel intensity measurement. The machine directly programs fully connected 797-spin Ising models with external fields at nearly 9-bit precision and operates at a constant iteration rate of $\sim 200$ Hz. By removing zero-valued product terms, the same architecture scales to sparse problems and experimentally solves a Max-cut instance on a 424,108-vertex M\"obius ladder graph. We also observe the phase transition of the Sherrington--Kirkpatrick model, demonstrating programmable optical simulation beyond low-rank couplings. These results establish amplitude modulation as a scalable route to programmable photonic Ising machines.
\end{abstract}

\maketitle

\textit{Introduction.}--- Combinatorial optimization problems are prevalent in industrial production and scientific research, such as circuit design \cite{brophy2014}, financial portfolio management \cite{soleymani2021}, and compressed sensing problems \cite{zheng2024,zheng2026,gao2025}, which aim to search a set of feasible solutions that satisfy constraints and optimize the value of the objective function within a large solution space. However, the traditional exhaustive searching algorithm exhibits poor/failed performance when the problem scale expands, as the dimensionality of its solution space increases exponentially. To efficiently solve such large-scale complex problems, a class of physics-inspired special-purpose computers (i.e., Ising machines) distinct from the von Neumann architecture has been proposed, such as coherent Ising machines (CIMs) \cite{honjo2021,inagaki2016}, atomic Ising machines \cite{kiraly2021}, and optoelectronic oscillators Ising machines \cite{wu2025}. The pending practical problems are formulated as minimization problems of the Ising Hamiltonian,
\begin{equation}
H_{\mathrm{Ising}}=-\sum_{i,j}J_{ij}\sigma_i\sigma_j-\sum_i h_i\sigma_i,
\label{eq:ising}
\end{equation}
where $H_{\mathrm{Ising}}$ is the Ising Hamiltonian, $J$ is the spin coupling matrix, $h$ is the external field vector, $\sigma$ is the spin vector with values in $\{-1,1\}$, and $i$ and $j$ are the corresponding indices.

\begin{figure*}[t]
\centering
\includegraphics[width=\textwidth]{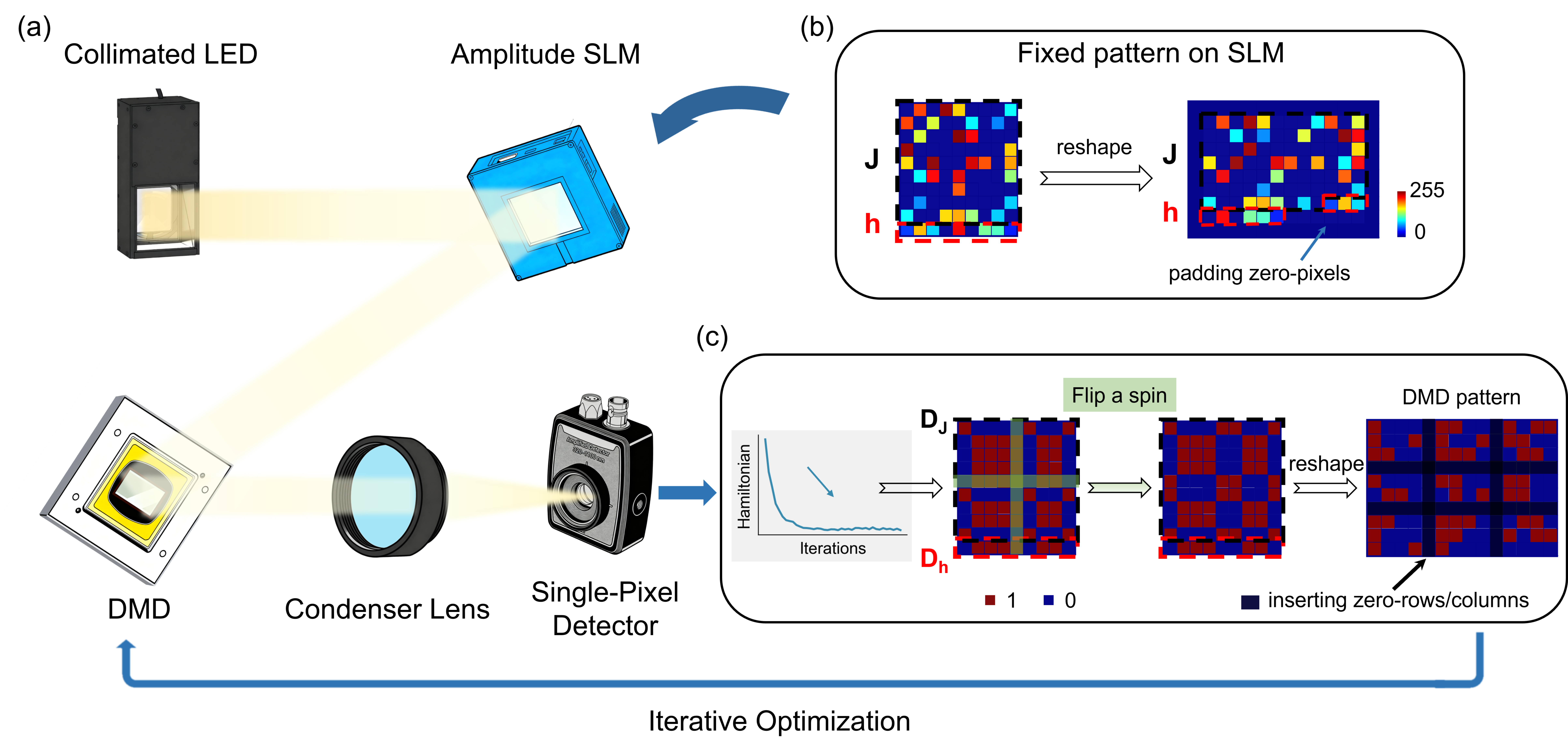}
\caption{Scheme and principle of the proposed AR-SPIM. (a) The experimental setup. (b) The fixed pattern, which is designed and reshaped by $|J|$ and $|h|$ (corresponding to the black and red dashed region) of the target Ising problem, is uploaded to the A-SLM. A collimated incoherent LED light illuminates the A-SLM. The modulated light, after scaling and filtering by the 4-$f$ system (not shown in the figure), illuminates the DMD. The reflected light is focused onto the single-pixel detector through a condenser lens. (c) The Metropolis-Hastings sampling procedure of the proposed spatial photonic Ising machine. The binary DMD pattern is designed and reshaped by $D_J$ and $D_h$ (corresponding to the black and red dashed region). The pixels of the A-SLM and DMD are carefully aligned with the insertion of zero rows and columns to reduce errors (\textbf{see Supplemental Material Secs. S1 and S2}). After the intensity measurement, a spin flip is randomly selected, followed by inverting the pixels at the corresponding coordinates on the DMD, and the intensity is measured again. This flipping-measurement process (\textbf{see Supplemental Material Sec. S3}) is repeated in each iteration, and the final spin configuration is intended to minimize the intensity.}
\label{fig:scheme}
\end{figure*}

The recently proposed spatial photonic Ising machine (SPIM) \cite{pierangeli2019,ouyang2024,sun2025,yu2024,olivieri2025,leonetti2021}, founded on the principle of laser diffraction, offers an efficient and scalable approach for implementing large-scale fully connected Ising models. By modulating the optical field with phase spatial light modulators (P-SLMs) and encoding the intensity at the detector center into the Ising Hamiltonian, the system evolves toward the optimal solution through feedback-based optical annealing. However, to accommodate external fields and high/full-rank spin-spin coupling matrices in practical applications, SPIM must be combined with approaches such as the auxiliary-spin-pair method \cite{sakellariou2025} and wavelength, spatial, or temporal division multiplexing \cite{luo2023,veraldi2025,yamashita2023}, at the cost of reduced time efficiency or a smaller number of spins.

In this letter, we propose and demonstrate an amplitude-only modulated, rank-free SPIM (AR-SPIM) that can directly incorporate external fields and rank-free spin couplings. AR-SPIM leverages Hadamard product decomposition to encode arbitrary Ising models onto an incoherent light field via a carefully aligned amplitude spatial light modulator (A-SLM) and digital micromirror device (DMD), avoiding the waste of spin and time resources, thereby enhancing the utilization efficiency of optical hardware. Compared with the P-SLMs-based SPIMs (60 Hz), our proposed AR-SPIM eliminates coherent interference aberrations while achieving nearly linear and 9-bit encoding precision, and delivers an acceleration of at least 3.3 times. We have successfully demonstrated the full programmability of AR-SPIM in experimentally solving combinatorial optimization problems and phase transition observations of the Sherrington-Kirkpatrick (SK) model. In comparison with SPE\cite{sakellariou2025} and other SPIM schemes, our proposed scheme has advantages of utilize all pixels on the SLM and DMD to encode the large-scale sparse Ising model, overcoming the limitations imposed by incoherent aberrations, which restricts the encoding to a small number pixels of the center region of the phase-only SLM. We have developed an adaptive strategy to increase the available spin counts for sparse Ising problems via removing zero-valued Hadamard product terms, and the AR-SPIM scores exceed the 90\%-best solution of the Max-cut problem for a 424,108-vertex M\"obius ladder graph. This approach has experimentally solved the maximum-scale full-rank Ising problem, surpassing the state-of-the-art of existing programmable SPIMs (\textbf{see Appendix A}), and would further inspire the development of universal, high-speed-precision Ising machines.

\begin{figure*}[t]
\centering
\includegraphics[width=\textwidth]{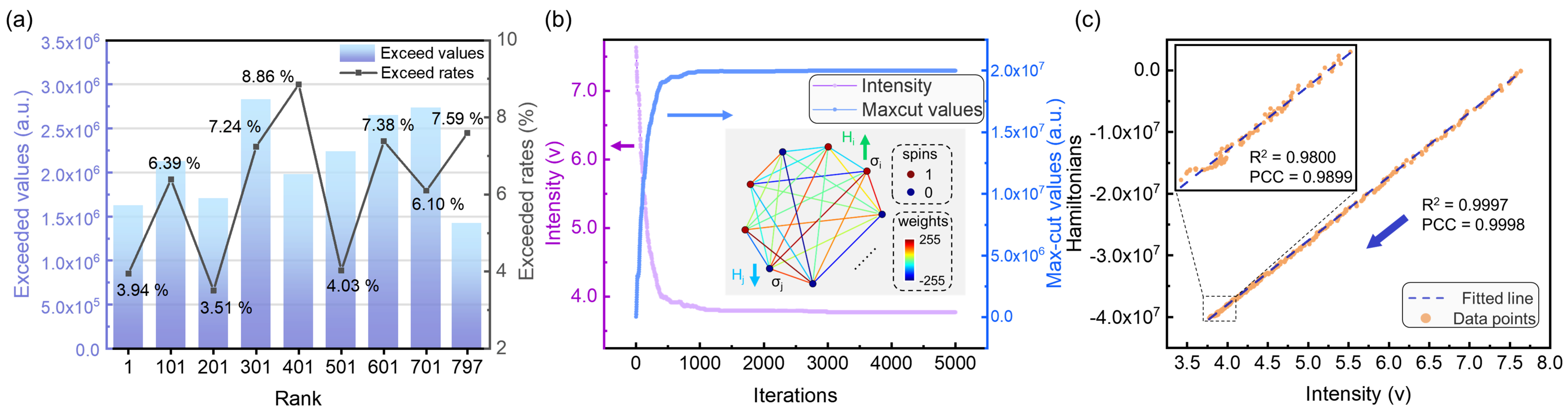}
\caption{Performance of AR-SPIM in solving the 797-spin arbitrary-weighted Max-cut problems with external field biases. (a) The quantitative comparison of the AR-SPIM scores and GW-SDP scores at different ranks (1, 100, 200, $\ldots$, 700, 797). The exceeded values (the colored bar charts) and exceeded rates (the gray points and line) are shown. (b) The measured intensities (the purple line) and calculated Max-cut values (the blue line) at various iterations of the full-rank problem. The schematic diagram of this problem is inserted in (b). (c) Quantitative linearity analysis for the case in (c), which shows the calculated Ising Hamiltonians versus the measured intensities. The orange dots represent the data points after removing duplicate data, and the purple dashed line is the linear fit to the data points. For global data points, the coefficient of determination $R^2$ for the linear fit is 0.9997, and the Pearson correlation coefficient (PCC) is 0.9998. In the region close to the reference ground state ($-4.02\times10^7$) (i.e., intensity $<4.2$ V), the $R^2$ is 0.9800, and the PCC is 0.9899. The purple arrow indicates the direction of optical annealing.}
\label{fig:maxcut797}
\end{figure*}

\textit{Physical Model.}--- Instead of eigenvalue decomposition, the arbitrary Ising Hamiltonian is re-formulated in the form of the summation of two Hadamard products in principle,
\begin{equation}
H_{\mathrm{Ising}}=\mathrm{sum}\left\{\left[|J|\odot\left(-S_J\odot\sigma\sigma^T\right)\right]\oplus\left[|h|\odot\left(-S_h\odot\sigma\right)\right]\right\},
\label{eq:hadamard}
\end{equation}
where $\mathrm{sum}$ denotes the matrix/vector summation function, $\oplus$ denotes the matrix/vector vertical concatenation function, and $\odot$ denotes the Hadamard product operation (i.e., element-wise multiplication of matrices/vectors). $|J|$ and $|h|$ are the absolute values matrix/vector of $J$ and $h$, with 8-bit precision, which can be combined as a fixed pattern and uploaded into the A-SLM. $S_J$ and $S_h$ are the sign functions that record the signs of elements of matrix $J$ and vector $h$, defined as
\begin{equation}
\left(S_J\right)_{ij}=\begin{cases}1, & J_{ij}\ge 0,\\ -1, & J_{ij}<0,\end{cases}\qquad
\left(S_h\right)_i=\begin{cases}1, & h_i\ge 0,\\ -1, & h_i<0.\end{cases}
\label{eq:sign}
\end{equation}
We set the elements with a value of $-1$ in $(-S_J\odot\sigma\sigma^T)$ and $(-S_h\odot\sigma)$ to 0. The binary patterns $D_J$ and $D_h$ are defined as
\begin{subequations}
\begin{align}
\left(D_J\right)_{ij}&=\begin{cases}1, & \left(-S_J\odot\sigma\sigma^T\right)_{ij}=1,\\ 0, & \left(-S_J\odot\sigma\sigma^T\right)_{ij}=-1,\end{cases}\label{eq:dj}\\
\left(D_h\right)_i&=\begin{cases}1, & \left(-S_h\odot\sigma\right)_i=1,\\ 0, & \left(-S_h\odot\sigma\right)_i=-1.\end{cases}\label{eq:dh}
\end{align}
\end{subequations}
As shown in Fig.~\ref{fig:scheme}, by reshaping and imposing the $D_J$ and $D_h$ patterns onto the DMD, the total intensity $I$ of the modulated incoherent light field is expressed as
\begin{equation}
I=\mathrm{sum}\left\{\left[|J|\odot D_J\right]\oplus\left[|h|\odot D_h\right]\right\}.
\label{eq:intensity}
\end{equation}
The process of summing matrices or vectors can be equivalent to collecting optical intensity, thereby enabling a quadratic speed enhancement in comparison to electronic computers (\textbf{see Supplemental Material Sec. S9}). The relationship between the Ising Hamiltonian $H_{\mathrm{Ising}}$ and $I$ can be expressed as
\begin{equation}
H_{\mathrm{Ising}}=2I-C,
\label{eq:linear}
\end{equation}
where $C=\mathrm{sum}\{|J|\oplus |h|\}$ is a constant of the specific target problem, and $H_{\mathrm{Ising}}$ and $I$ exhibit a linear relationship. Therefore, the scheme of AR-SPIM is programmed into an amplitude-modulated light field, and we can search the ground state of the Ising problem through optimizing $I$ instead of $H_{\mathrm{Ising}}$ by the proposed AR-SPIM, as shown in Fig.~\ref{fig:scheme}. The available spin counts for undirected graphs can reach $\frac{\sqrt{2nm}}{1-\gamma}$, with a spatial complexity of $O(E)$, where $\gamma$ represents the proportion of zero elements, $E$ represents the edges of the graph, and the alignment region is $n\times m$ pixel-pairs.

\textit{Max-cut solving.}--- To demonstrate the ability of AR-SPIM for solving combinatorial optimization problems, we conduct experiments for two types of Max-cut problems (\textbf{see Supplemental Material Sec. S5}), including 797-spin arbitrary-weighted graphs with external field bias, and a sparse, large-scale M\"obius ladder graph. We have also solved the maximum independent set (MIS) problems (\textbf{see Supplemental Material Sec. S6}). In the arbitrary 797-spin Max-cut problem experiments, nine spin-spin coupling matrices (i.e., $J$, with a size of $797\times797$) with distinct ranks (1, 100, 200, $\ldots$, 700, 797) are generated with arbitrary values at near 9-bit precision (from $-255$ to 255). Each problem is conducted with a random bias vector (i.e., $h$, with a size of $1\times797$), which is defined with a 60\% probability of $h_i=2$ and a 40\% probability of $h_i=-1$. The initial spin configuration is consistently defined as an all-ones $1\times797$ vector. For each case, two aligned pixel pairs are grouped into a super pixel pair to encode a Hadamard-product term. The processed and reshaped $|J|$, $|h|$, $D_J$ and $D_h$, obtained through Eqs.~\eqref{eq:hadamard}--\eqref{eq:dh}, are programmed onto the A-SLM and DMD, respectively. Feedback-based optical annealing is subsequently performed on the AR-SPIM to search for the score $G_{\mathrm{exp}}$. As the solution is as large as $2^{797}$ possibilities, which can hardly be exhausted, we utilize the GW-SDP algorithm to obtain the reference scores $G_{\mathrm{ref}}$ for each case, which can guarantee an expected value of at least 87.8\% of the best solution \cite{inagaki2016,goemans1995}. We define the exceeded values $G_{\mathrm{exp}}-G_{\mathrm{ref}}$, and the exceeded rate $(G_{\mathrm{exp}}-G_{\mathrm{ref}})/G_{\mathrm{ref}}$. As illustrated in Fig.~\ref{fig:maxcut797}(a), the Max-cut scores obtained by AR-SPIM are consistently higher than those of GW-SDP, with increments ranging from 3.51\% to 8.86\%. This result guarantees the accuracy of AR-SPIM in solving any-rank combinatorial optimization problems. Figure~\ref{fig:maxcut797}(b) presents the annealing process of full-rank problems with iterations. The relationship between Hamiltonian and light intensity is displayed in Fig.~\ref{fig:maxcut797}(c), where both global and local scales exhibit a reliable linear mapping relationship matching Eq.~\eqref{eq:linear}.

\begin{figure*}[t]
\centering
\includegraphics[width=\textwidth]{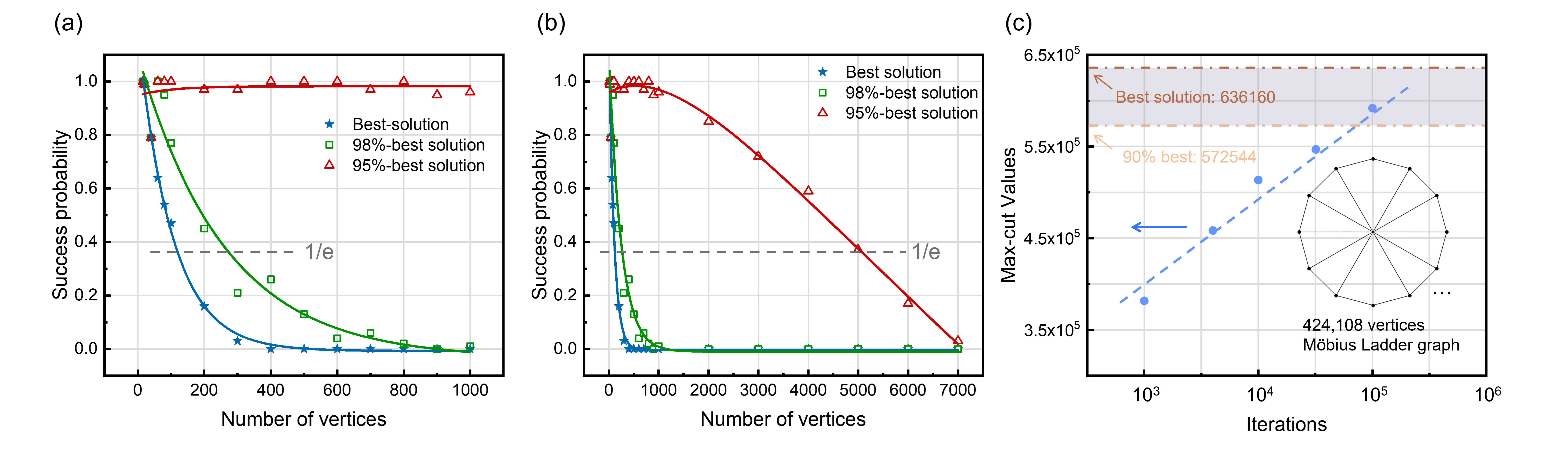}
\caption{Performance of AR-SPIM for the Max-cut problem on M\"obius ladder graphs of different sizes. (a),(b) The success probabilities of AR-SPIM after 20,000 iterations in achieving the best solution (the red triangles, i.e., ground truth), the 98\%-best solution (the green squares), and the 95\%-best solution (the blue stars) for the Max-cut problem on M\"obius ladder graphs from 0 to 1000 vertices, and from 0 to 7000 vertices, respectively. (c) The performance of AR-SPIM for the Max-cut problem on 424,108 M\"obius ladder graphs. The Max-cut score of AR-SPIM exceeds the 90\%-best solution before $10^6$ iterations.}
\label{fig:mobius}
\end{figure*}

 We quantitatively demonstrate the success probability of AR-SPIM for the Max-cut problems for M\"obius ladder graphs from 16 to 7000 vertices, which have theoretical closed-form solutions \cite{inagaki2016}. For each problem, we perform 100 runs with random initial spin configurations, and each run performs 20,000 iterations. Each vertex of M\"obius ladder graphs is connected only to its two neighbors and its centrally symmetric vertex, resulting in a coupling matrix with a lot of zero-valued elements. By removing these terms and loading the remaining non-zero elements and their corresponding terms onto the AR-SPIM, we can encode larger-scale graphs. AR-SPIM enables M\"obius ladder graphs with up to 424,108 vertices through employing an aligned pixel-pair to encode a Hadamard product term. We statistically analyze the success probabilities of the AR-SPIM in solving problems with different vertices for reaching the ground truth (i.e., best-solution), 98\% of the ground truth (i.e., 98\%-best solution), and 95\% of the ground truth (i.e., 95\%-best solution). The results for 16--1000 vertices and 16--7000 vertices are presented in Figs.~\ref{fig:mobius}(a) and \ref{fig:mobius}(b), respectively. As the scales increase, the success probabilities decrease. The success probability of achieving the ground truth and the 98\%-best solution decreases to $1/e$ ($\sim36.78\%$) when the number of vertices is approximately 120 and 280, respectively. The success probability of achieving the 95\%-best solution is near 100\% when the number of vertices ranges from 16 to 1000, and decreases to $1/e$ at approximately 5000 vertices. The detailed histograms are shown in Supplemental Material Sec. S7. We further demonstrate the evolving process of AR-SPIM for 424,108 vertices, as shown in Fig.~\ref{fig:mobius}(c). The AR-SPIM score increases with the iterations, and exceeds 90\%-best solution before $10^6$ iterations. The speed of AR-SPIM in addressing these issues does not increase with the number of spin (\textbf{see Appendix B}).

\begin{figure*}[t]
\centering
\includegraphics[width=\textwidth]{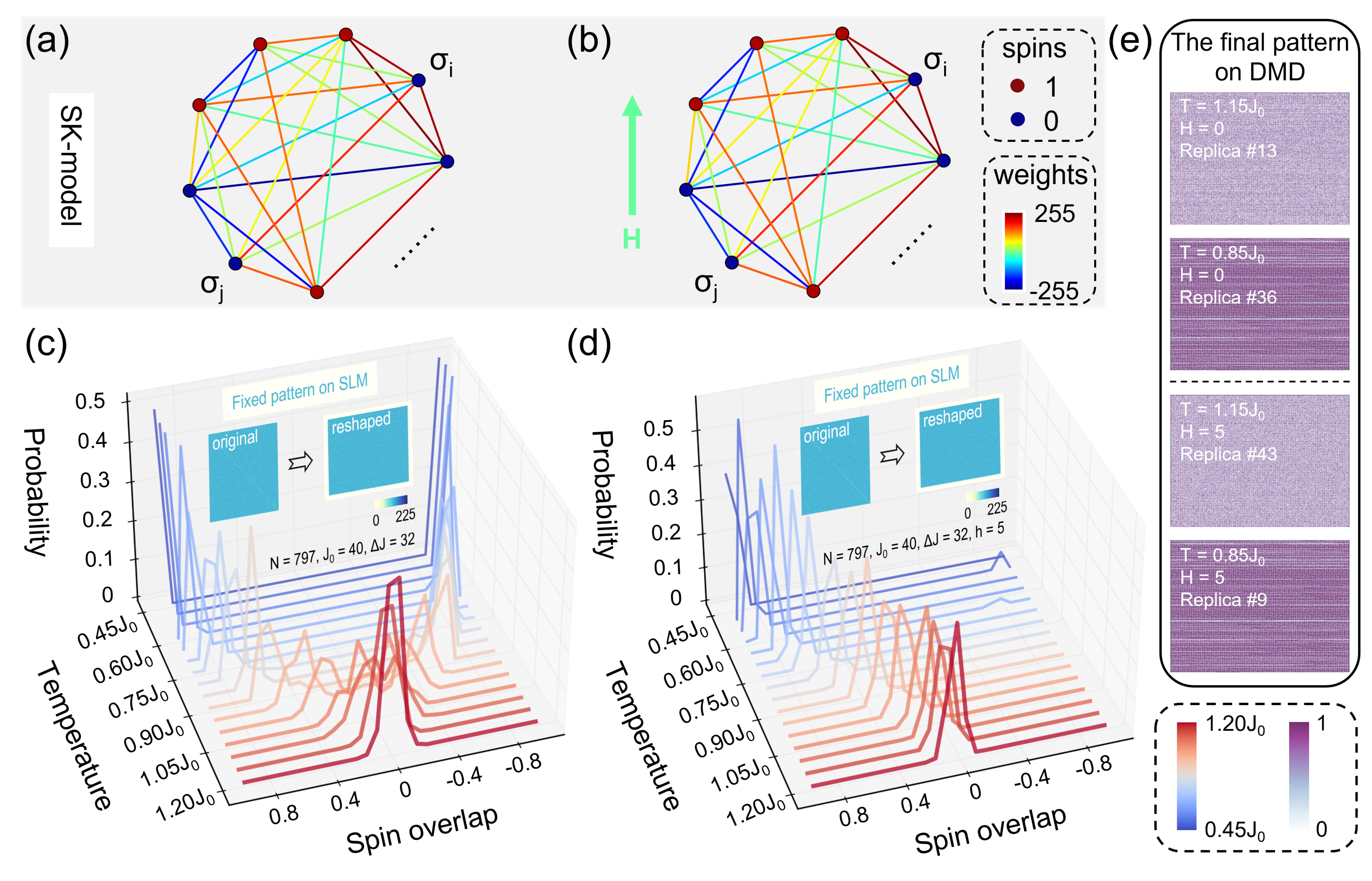}
\caption{Demonstration of AR-SPIM for simulating phase transitions in the SK model. (a) Schematic for SK model. (b) Schematic for the SK model with a uniform external magnetic field. (c),(d) The results of the three-dimensional (3D) phase diagram (probability-temperature-spin overlap) corresponding to the scenario in (a) and (b), respectively. Correspondingly loaded patterns on the A-SLM, which encode spin coupling and external magnetic field, are presented in (c) and (d), respectively. (e) The final evolved patterns on the DMD after 5000 iterations by AR-SPIM for each selected replica.}
\label{fig:sk}
\end{figure*}

\textit{Phase transition of SK Model.}--- We demonstrate the phase transitions of the SK model. As shown in Fig.~\ref{fig:sk}(a), we set $N=797$, $J_0=40$, and $\Delta J=32$. The spin-spin coupling matrix $J$ is randomly generated, which is presented in Fig.~\ref{fig:sk}(c). In accordance with the mean-field theory \cite{nishimori2001}, when $J_0/\Delta J>1$, the SK model undergoes a phase transition from the ferromagnetic (FM) phase to the paramagnetic (PM) phase as the temperature increases from low to high. The critical transition temperature, denoted by $k_BT_c$, is equal to $J_0$, where $k_B$ is the Boltzmann constant. We experimentally utilize the replica method to simulate the 3D phase diagram of the SK model (\textbf{see Supplemental Material Sec. S8}). We generate 50 initial spin replicas composed of $-1$ and 1 randomly. At each temperature, each replica evolves toward the equilibrium state after 5000 iterations through AR-SPIM. We calculate and statistically analyze the spin overlap \cite{nishimori2001} across 16 different temperatures (from $0.45J_0$ to $1.2J_0$). As presented in Fig.~\ref{fig:sk}(c), at high temperature ($>J_0$), the spin overlap value clusters around zero, indicating that the evolved spin configurations of the 50 replicas are randomly distributed, which demonstrates the PM phase of the SK model. At low temperature ($<J_0$), the probability distribution of spin overlap manifests a bimodal characteristic. As the temperature is reduced, the spins within the same replica align simultaneously in the upward (+1) or downward direction ($-1$), with the resultant magnetization intensity tending towards +1 or $-1$, which demonstrates the FM phase. As shown in Figs.~\ref{fig:sk}(b) and \ref{fig:sk}(d), upon applying a uniform external magnetic field ($h=5$) to the SK model, the system's symmetry is broken, eliminating the degenerate ferromagnetic ground states and inducing all spins to tend to align with the direction of the external magnetic field. Thus, the spin overlap distribution transforms from a bimodal to a unimodal one. This phenomenon also reflects that the application of an external magnetic field can regulate the magnetization behavior.

\textit{Conclusion and Outlook.}--- In conclusion, we have proposed the AR-SPIM for encoding and solving the Ising models with rank-free spin-spin coupling and external fields, which requires only once programming to map high/full-rank Ising models and inspires training of large-scale artificial intelligence. We present the highly accurate, stable, and high-speed characteristics of AR-SPIM through the efficient tackling of arbitrary Max-cut problems and observing the phase transitions in the SK model. We also offer an adaptive strategy to increase the available spin counts for the sparse Ising problem. We have also supplemented the maximum spin count under alignment tolerance, the time analysis, and the long-term operation tests to demonstrate the advantages and stability of AR-SPIM (\textbf{see Supplemental Material Secs. S4, S9, and S10}).

Moreover, the advantages of AR-SPIM are based on the Hadamard product decomposition and incoherent amplitude modulation, which demand high-precision alignment of A-SLM and DMD. The fusion of employed devices with higher resolutions and a higher rate would further enhance the performance of AR-SPIM. More robust techniques for detecting weaker light intensity (such as single-photon detection and quantum precision measurements) should be integrated to ensure analogue signals are noiseless and accurately identify the Ising model’s Hamiltonians while searching close to the ground state. Ultrahigh-bandwidth micro-LED chips' modulation schemes are expected to boost the performance of AR-SPIM to the GHz\cite{lu2022dennard,parbrook2021micro,liu2026fast,yahav2021multi}.

\begin{acknowledgments}
This work was supported by the National Key R\&D Program of China (No. 2025YFF0515504), the National Natural Science Foundation of China (Nos. 62401359 and 62471289), the State Key Laboratory of Photonics and Communications, the Quantum Science and Technology---National Science and Technology Major Project (No. 2021ZD0300703), and the Shanghai Municipal Science and Technology Major Project (No. 2019SHZDZX01).
\end{acknowledgments}

\appendix

\section{Comparison of the experimentally demonstrated rank and spin
counts across fully programmable spatial photonic Ising machines
(SPIMs).}

\begin{figure*}[t]
\centering
\includegraphics[width=\textwidth]{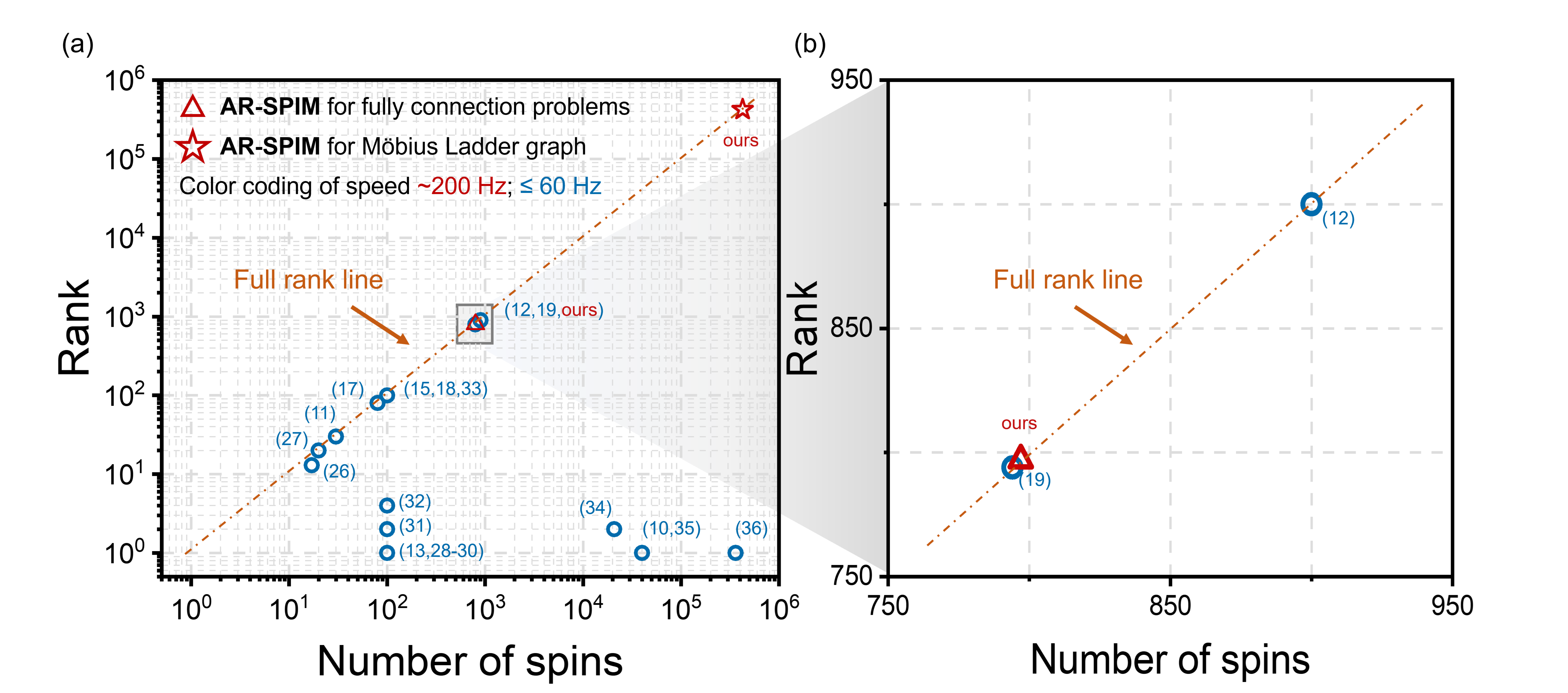}
\caption{Comparison of the experimentally demonstrated rank and number of spins across fully programmable spatial photonic Ising machines (SPIMs)\cite{pierangeli2019,ouyang2024,sun2025,yu2024,sakellariou2025,luo2023,veraldi2025,yamashita2023,sakabe2023spatial,wang2024general,fang2021experimental,pierangeli2020adiabatic,pierangeli2020noise,sun2022quadrature,shimomura2025parallel,feng2025spin,ye2025smoothed,huang2021antiferromagnetic,yao2025enhanced}. The cycles represent existing SPIMs. The triangle represents the maximum rank and spin count of AR-SPIM for fully connected problems. The star represents the maximum rank and spin count of AR-SPIM for the M\"obius ladder graph. Blue represents an iteration rate of approximately 200 Hz, and red represents an iteration rate of less than 60 Hz. The orange dashed line represents the full rank line. (a) The spin number and rank range from 1 to 10$^6$. (b) The spin number and rank range from 750 to 950. The numerals in parentheses correspond to the reference numbers}
\label{fig:com}
\end{figure*}

As shown in Fig. 5, the proposed AR-SPIM achieves the highest iteration rate ($\sim$ 200 Hz), surpassing all existing SPIMs based on
phase-only spatial light modulators ($<$ 60 Hz). The AR-SPIM has experimentally verified the successful experimental solution to the current largest-scale problem (i.e., the 424,108 vertices M\"obius Ladder graph with full rank). For the full-rank fully connected Ising problem, the maximum number of spins AR-SPIM can encode is on par with the existing state-of-the-art (SOTA)\cite{sun2025} experimentally, while its iteration speed is at least 3.3 times higher than that of the SOTA.

\section{The O(1) time scale for 16 - 424,108 vertices Max-cut problems of M\"obius ladder graphs}

\begin{figure*}[t]
\centering
\includegraphics[width=0.6\textwidth]{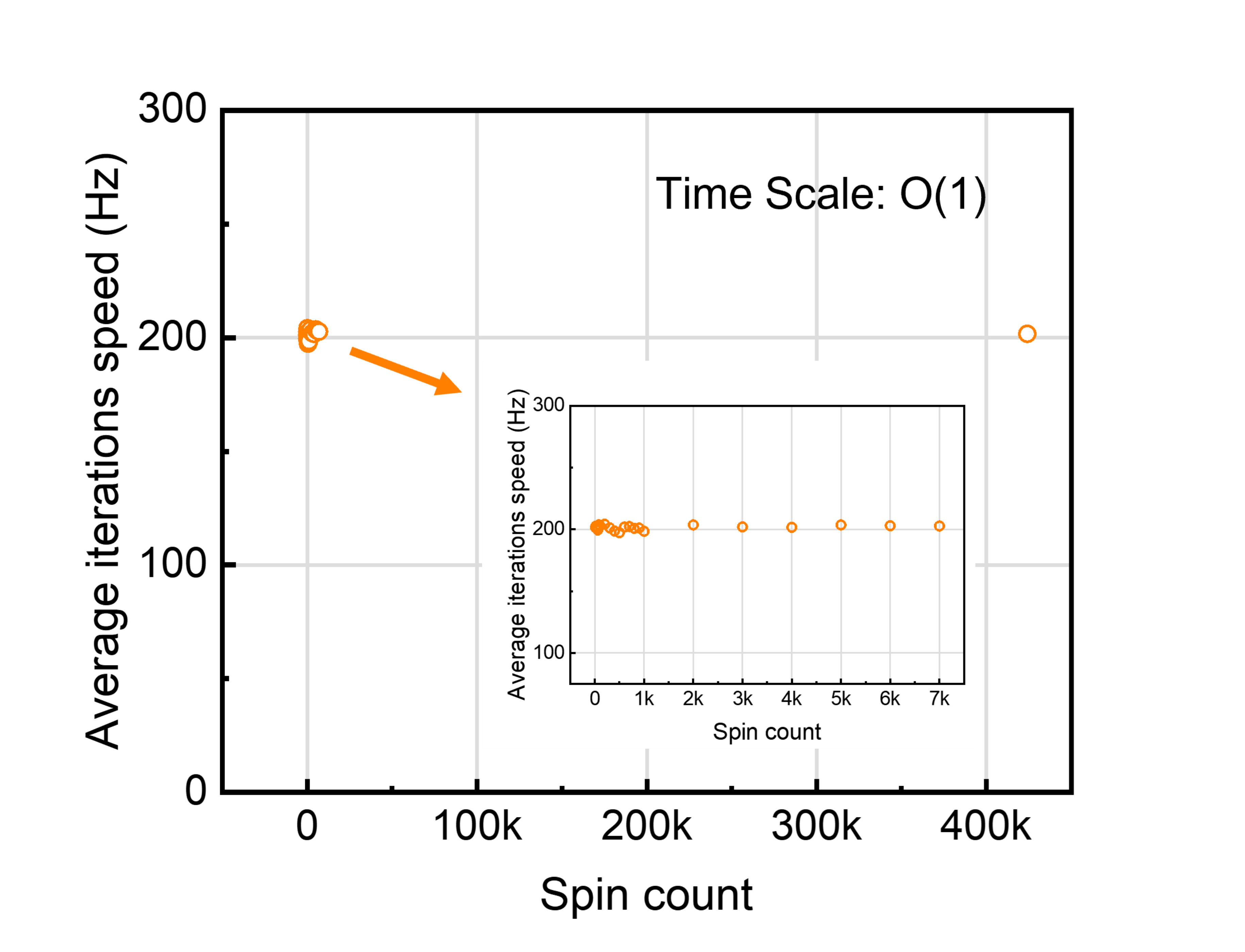}
\caption{The average iteration speed of AR-SPIM for the Max-cut problem on M\"obius ladder graphs of different sizes (orange
cycles). The time scale is O(1) with extant experimental results.}
\label{fig:o1}
\end{figure*}

We further provide experimental evidence that the iteration speed of AR-SPIM is O(1) with respect to spin counts. We record and demonstrate the iteration speed of AR-SPIM in solving the Max-cut problem for large-scale M\"obius ladder graphs (as shown in Fig. 3 of the main text). It can be observed that as the number of vertices increases from 16 to 424,108, the average iterations per second of AR-SPIM remains stable at approximately 200 Hz. This indicates that the iteration speed of AR-SPIM does not increase with the problem scale. Instead, as the problem scale increases, the speed advantage of AR-SPIM becomes more pronounced, as illustrated in Fig. 6.

\nocite{*}

\bibliography{references}

\end{document}